# Temperature Micro-Raman Study of Lysozyme Crystals

Vasylkiv Yu.[1], Polomska M.[2], Nastishin Yu.[1] and Vlokh R.[1]

[1] Institute of Physical Optics, 23 Dragomanov St., 79005 Lviv, Ukraine, e-mail: vlokh@ifo.lviv.ua

[2] Institute of Molecular Physics of the Polish Academy of Sciences, 17 M. Smoluchowskiego St., 60179 Poznan, Poland

**Abstract**

We have performed for the first time micro-Raman studies of lysozyme crystals for the temperatures lower than the room one (in the range $282-300\,\text{K}$). Anomalous temperature dependence of the positions of Raman scattering bands has been detected and discussed, which corresponds to the crystalline lysozyme structure in the vicinity of $T=285\,\text{K}$.

**Introduction**

Successes in decoding of genomes for a number of living organisms have demonstrated a power of molecular biology, when equipped by physical and chemical characterization techniques. High effort of biologists is now focused on different studies of proteins. Single crystals composed of biological molecules (referred to as biocrystals further on) represent a unique possibility for structural characterization of biological molecules. Progress in a new field of biology, proteomics, is largely dependent on availability of protein samples. In the majority cases, the studies have been performed on single biocrystals. Structural phase transitions in crystals remain a top-ranking problem in the condensed matter physics for many years. Interest to the problem of phase transitions in single biocrystals is twofold. First, it is related to the functional role of proteins in the living organisms. Structural transformations in single biocrystals can shed light on physical properties of protein molecules, which can be at work when explaining molecular mechanisms of diseases accompanied with pathology in the structure and biological activity of protein molecules. Second, since the proteins are macromolecules that form crystalline lattices with giant elementary cells, which in addition are significantly hydrated, it would be important to probe the typical approaches developed for the crystal physics of these objects. While there is a big progress in understanding of biochemistry of proteins, their physical properties remain to be an open field.

Lysozyme crystals grown and existing in their mother solution are among the most stable, relatively easily grown biocrystals. Probably, because of this, they are the most studied, though not enough when compare with the conventional inorganic crystals. According to the literature [1,2], the lysozyme crystals can be grown in two crystalline modifications, the orthorhombic ($P2_12_12_1$, when grown at the temperatures several degrees above the room temperature or higher) and tetragonal ( $P4_32_12$ , when grown at the room temperatures and below). A phase transition at the temperatures some Kelvin degrees above the room one has been suspected in references [1,2], basing on the observations that the tetragonal crystals dissolve and then grow in orthorhombic modification on heating. Encouraged by the fact documented in [3] that, literally citing, "the sample denaturates" above 307 K, the authors of Ref. [4] have studied Brillouin light scattering and have indeed found clearly expressed anomalous behaviour in the vicinity of this temperature, which for this reason was interpreted as an evidence of some phase transition between different crystalline modifications (though without any specification about the nature or symmetry of the crystalline phases). Intrigued by this challenging report, we have performed polarization microscopy observations aimed to detect textural transformations corresponding to this transition but we have found instead that the tetragonal lysozyme crystals just dissolve in their mother solution above 307 K [5,6]. To check whether the crystals are unstable above 307 K , being out of the bulk of their mother solution, we have carried out similar observations for crystal specimens, which appeared out of the mother solution after drying of the cell. We do not exclude that such specimens are wetted by a solution film, but it has been seen that, even if the liquid film coats the crystal, it is not thick and these conditions are similar to those described in Ref. [4]. It is worth adding that total drying is fatal for the crystal. As a result, we have concluded that above 307 K tetragonal lysozyme crystals are unstable and, thus, the anomaly reported in [4] displays in fact the limit of stability of the crystal, which has been termed in [3] as "sample denaturation".

In the same vein, we are still aware of the problem that our recent report about the phase transition at T=284 K in lysozyme crystals [7] has been based purely on polarization microscopy observations, which are difficult for deducing the exact structural information and, thus, need to be inspected by means of structural characterization techniques, providing direct measurements of structural crystalline parameters. Raman scattering of light is among such the techniques. Because of tiny sizes of specimens under study, which are usual for biocrystals, we deal with a modified version of the set-up further called as a micro-Raman set-up. There is an essential contribution to this field in the literature on lysozyme crystals (see the works [8-12]). However, all the available data concern the temperatures above the ambient temperature. The very aim of this paper is to present results of the micro-Raman studies covering the temperature range of 282-300 K .

**Experimental**

The NIR FT Raman spectra were measured by using a Bruker IFS66FRA106 Raman spectrometer equipped with a Bruker Ramanscope. The samples were excited with $\lambda = 1064$ nm diode pumped with a Nd:YAG laser. The spectral resolution was not worse than $2 \, \text{cm}^{-1}$ . The Lincam heating/cooling stage was used for the temperature control ( $278 - 308 \, \text{K}$ ). Each of the measured Raman spectrum resulted from the computer averaging of 1500 scans. These spectra were fitted using the standard Origin software (OriginLab Corporation).

The lysozyme crystals were grown following the receipt described in our recent paper [6]. For the Raman scattering studies, we use samples with the average size of $0.1\times0.1\times0.1\text{mm}^3$ in flat glass capillary. The capillary has been filled by the lysozyme solution (or the mother growth liquid). The spectra for the glass, the lysozyme mother liquid and the pure lysozyme crystals were recorded in order to facilitate identification of the Raman data corresponding to just the lysozyme crystals. These spectra are presented in Fig.1. The term "pure crystal" is reserved further on for the lysozyme crystal, which has been removed from its mother solution and placed onto a substrate.

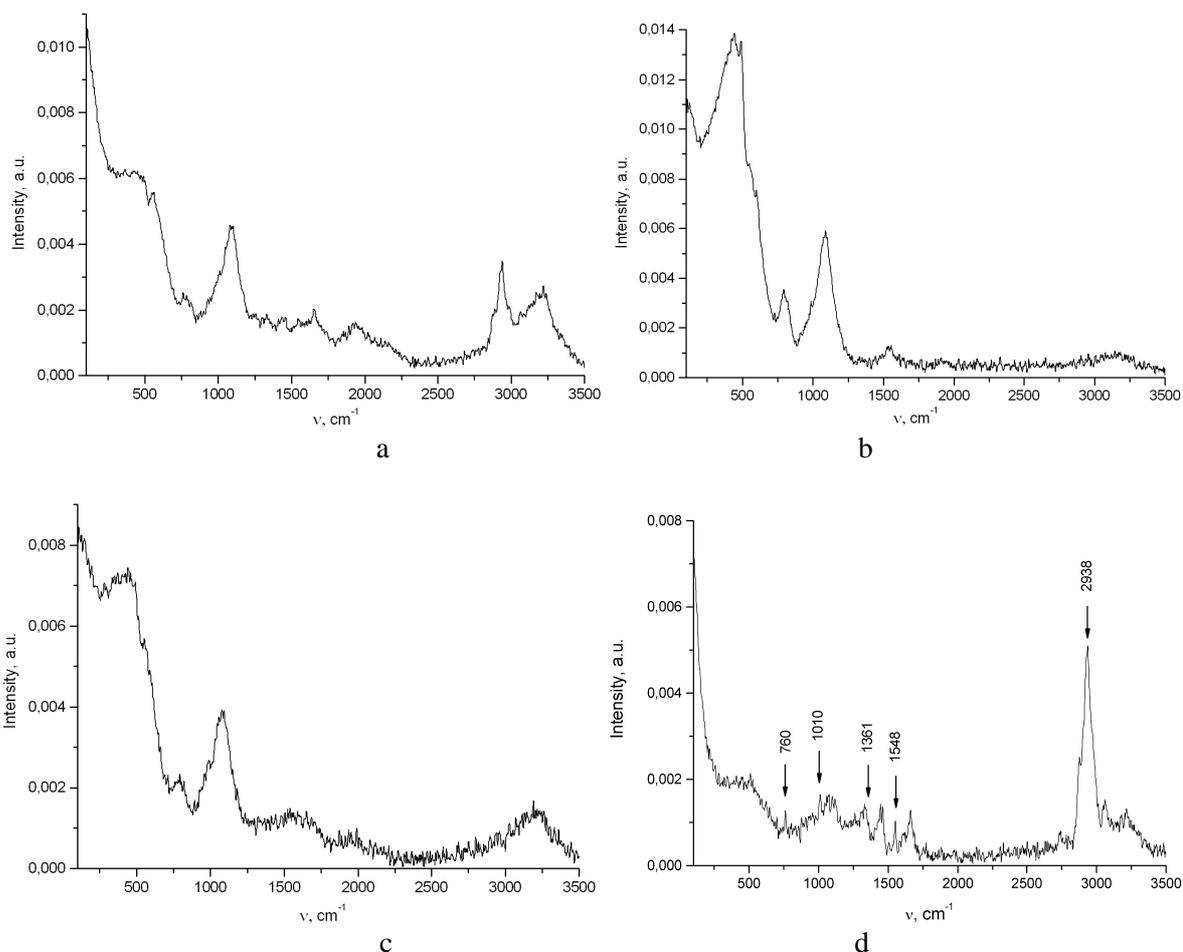

Figure 1. Raman spectra for the lysozyme crystal placed into flat cell with the lysozyme solution (a), flat glass cell (b), lysozyme solution in the glass cell (c) and pure lysozyme crystal (d) at the room temperature.

**Results and discussion**

Referring to the available literature data [8-12], one can identify at least five Raman bands in Fig. 1d, corresponding to the lysozyme molecules and centred at $760\text{cm}^{-1}$ (it is assigned to symmetric benzene/pyrrole in-phase breathing in tryptophan and symmetric ring breathing in tyrosine and phenylalanine), $1010\text{cm}^{-1}$ (symmetric benzene/pyrrole out-of-phase breathing in tryptophan), $1361\text{cm}^{-1}$ ($\delta(CH_2)$ and pyrrole ring vibrations in tryptophan), $1548\text{ cm}^{-1}$ (symmetric phenyl ring mode in tryptophan) and $2938\text{cm}^{-1}$ (C–H stretching band in $\nu(CH_2)$) [12].

It follows from the analysis of (Fig.1d) that the crystal lattice modes are presented exclusively by one set of the constitutive Gaussian bands around the frequency of $2900$ cm$^{-1}$, while the bands positioned in the range $750-1750$ cm$^{-1}$ are found also in the spectra for the lysozyme solution. Below we analyze the temperature behaviour of the constitutive Raman bands located around $2900$ cm$^{-1}$, keeping in mind that the very these bunds can point to the symmetry change at the crystalline phase transition.

We first analyze the temperature behaviour of the scattered intensity and then apply deconvolution of the spectra into constitutive Gaussian peaks. Three plots referred to 297, 282 and 278K display a more or less clearly expressed lowering of the scattering intensity for lower temperatures, while the plot corresponding to T=284K violates this tendency, crossing the other plots. While averaging through the spectral region shown in Fig. 2, the scattered intensity weakly changes with temperature in the high-temperature region of 13K above T=284K, though it decreases significantly below this temperature. Moreover, in the spectral region $2912 \div 2922$ cm$^{-1}$ the plots for 297K and 284K practically coincide but the intensity drops below T=284K. This fact might be related to a drastic increase in the diffusive light scattering occurring in the region below T=285K (see our previous papers [6,7]). It is caused by the formation of lysozyme aggregates, due to which this point might be termed as a "cloud point" [13].

Deconvolution of the spectra gives at least seven constitutive Gaussian peaks (see Fig. 3). The temperature dependence of the frequency of the most intensive peak is presented in Fig. 4. This dependence is anomalous and exhibits a minimum in the vicinity of 284K. Notice that this temperature is close to the temperature 285K identified as a special temperature point. According to our polarization microscopy observations [6,7] the latter could serve as a sign of phase transition in the lysozyme crystals.

In the pure lysozyme crystals we find a different situation: four constitutive Gaussian peaks obtained above 293K split into seven peaks below 282K (see Fig. 5). Again, this splitting can be apparent due to the increase in diffuse scattering of light occurred below 285K and worsening of the accuracy of Raman spectra recording. It is remarkable that the behaviour of the bands positioned in the spectral range of $750-1750$ cm$^{-1}$ (Fig. 6), corresponding to the lysozyme solution, is similar to the behaviour of pure crystalline lysozyme (compare to the data presented in Fig. 4). This might be due to the fact that the crystal is hydrated and contains a liquid solution inside of it. Moreover, the temperature shift of the band corresponding to the frequency $1361$ cm$^{-1}$ is larger than the absolute experimental error for the Raman spectra. The temperature positions of these frequency minima are slightly higher than those obtained for $2938$ cm$^{-1}$ (T=293K). Nonetheless, accounting that only three experimental temperature points are available, we admit that the exact positions of the minima need additional specification.

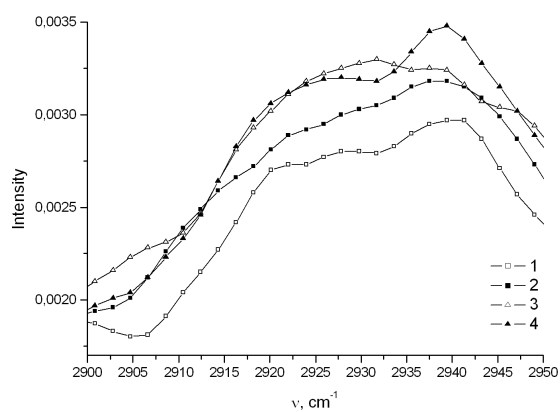

Figure 2. Raman spectra of the lysozyme crystal placed into flat cell with the mother liquid at 278 K (1), 282 K (2), 284 K (3) and 297 K (4). The spectral range under study is 2900-2950 cm$^{-1}$.

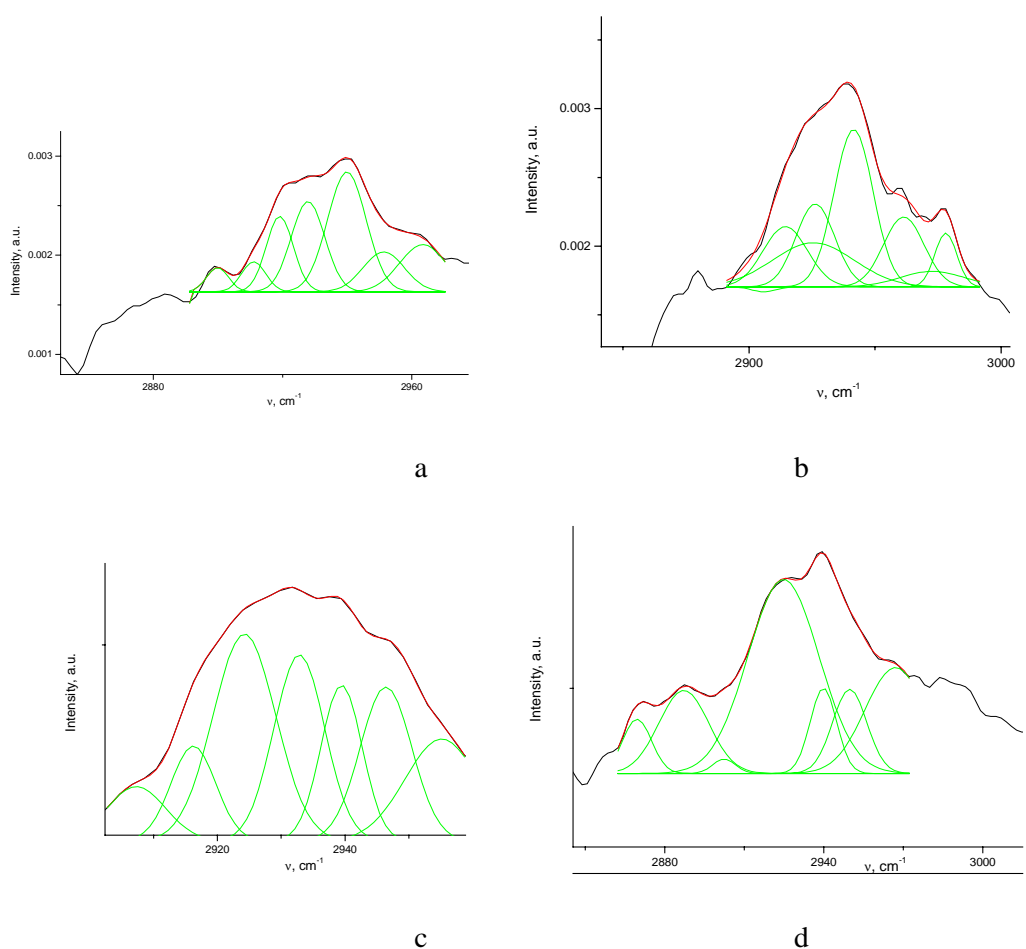

Figure 3. Fitting of Raman spectra of the lysozyme crystals placed in planar cell with the mother liquid (a – T=278 K, b – T=282 K, c – T=285 K and d – T=297 K).

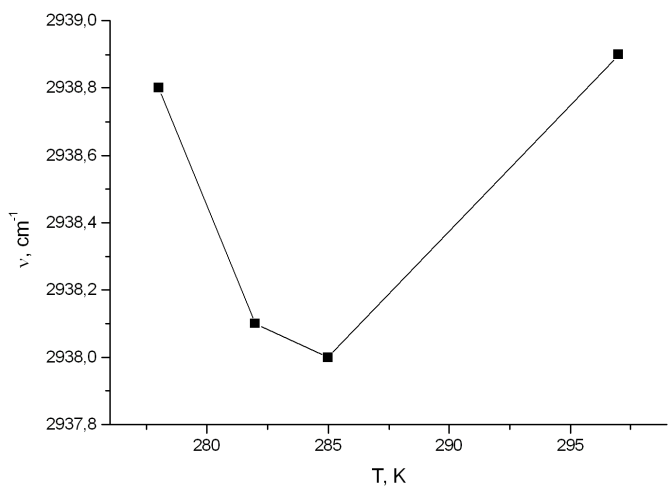

Figure 4. Temperature dependence of Raman peak frequency for the lysozyme crystals placed in the mother liquid at $2938\,\text{cm}^{-1}$.

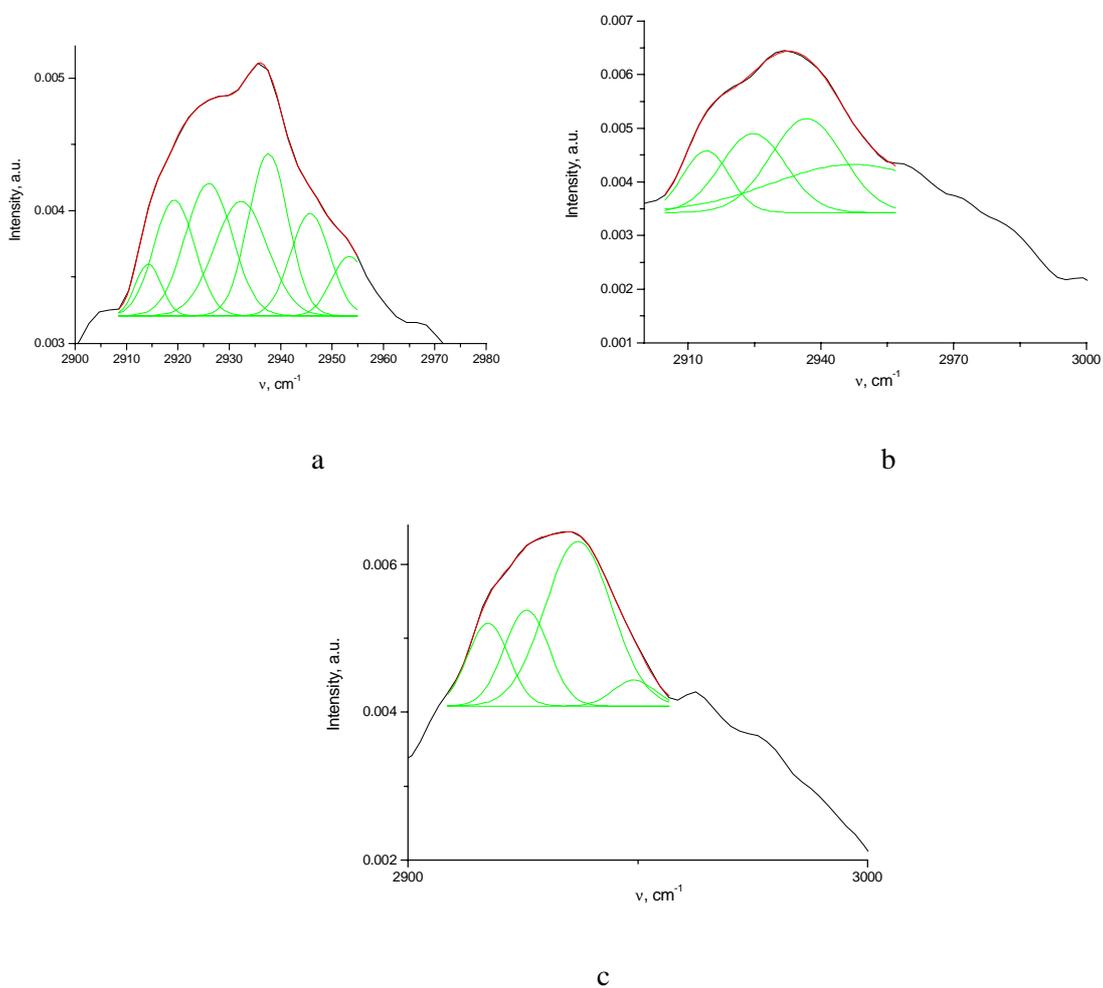

Figure 5. Fitting of pure lysozyme crystals Raman spectra (a – T=282 K, b – T=293 K, c– T=308 K)

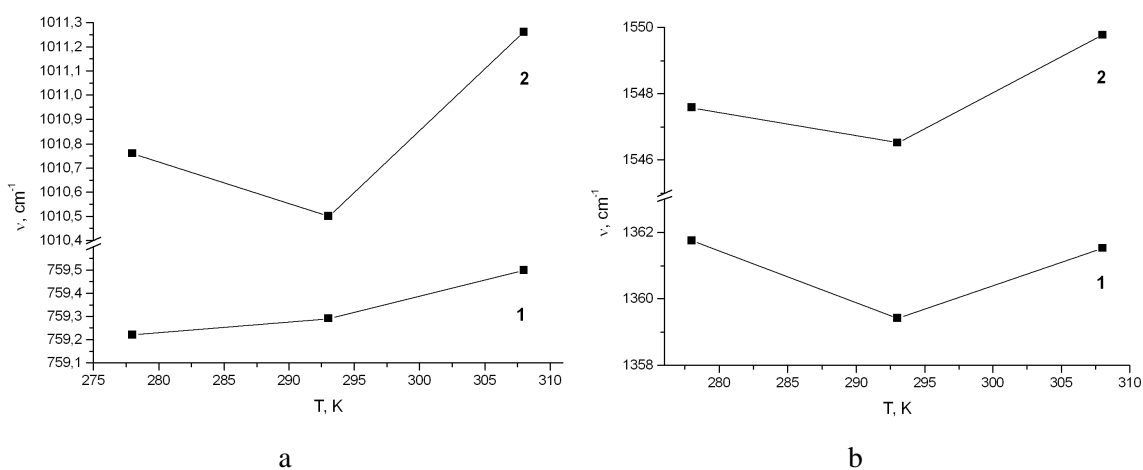

Figure 6. Temperature dependence of Raman peak frequency for the pure lysozyme crystals: (a) – at 759 cm$^{-1}$ (curve 1) and 1010 cm$^{-1}$ (curve 2) and (b) – at 1361 cm$^{-1}$ (1) and 1548 cm$^{-1}$ (2).

**Conclusions**

We have studied temperature behaviour of the structure of lysozyme crystals using the micro-Raman technique. It has been found that the temperature dependence of positions of all the Raman bands, which correspond to the crystalline lysozyme, exhibits minima in the vicinity of $T \approx 285\,K$. We have revealed that the four constitutive peaks obtained above $T=284\,K$ are split into seven peaks below this temperature. This could be related with the aggregation of lysozyme molecules taking place at $T \approx 285\,K$. In addition, we have found that the bands corresponding to the pure crystals also suffer anomalous behaviour at this temperature. Thus, one can suppose that the aggregation of molecules in the lysozyme solution can be responsible for rearrangement of lysozyme molecules in the crystal lattice. However, since a disappearance of any mode is not detected, it is not evident whether such a lattice modification is accompanied with some symmetry changes. As a consequence, the data presented above is not enough to justify the structural phase transition occurring in the lysozyme crystals at $T=284\,K$.

**Acknowledgement**

The authors[1] acknowledge financial support of this study from the Ministry of Education and Science of Ukraine (the Project N0106U000617).

**References**

1. Jolles P and Berthou J, 1972. High temperature crystallization of lysozyme: an example of phase transition. FEBS Lett. **23**: 22-23.
2. Berthou J and Jolles P, 1974. A phase transition in protein crystal: the example of hen lysozyme. Biochim. et Biophys. Acta. **336**: 222-227.
3. Kobayashi J, Asahi T, Sakurai M, Kagomiya I, Asai H and Asami H, 1998. The optical activity of lysozyme crystals. Acta Cryst. **A54**, 581-590.

4. Svanidze AV, Lushnikov SG and Seiji Kojima, 2006. Anomalous temperature behavior of hypersonic acoustic phonons in a lysozyme crystal. JETP Lett. **84**: 551-555.
5. Vasylkiv Yu, Nastishin Yu and Vlokh R, 2007. On the problem of phase transitions in lysozyme crystals. Ukr. J. Phys. Opt. **8**: 83-87.
6. Teslyuk I, Vasylkiv Yu, Nastishin Yu and Vlokh R, 2007. Structural phase transition in lysozyme crystals. Ferroelectrics. **346**: 49-55.
7. Teslyuk I, Nastishin Yu and Vlokh R, 2004. Structural phase transition in lysozyme single crystals. Ukr. J. Phys. Opt. **5**: 118-122.
8. Blanch EW, Hecht L and Barron LD. 2003. Vibrational Raman optical activity of proteins, nucleic acids, and viruses. Methods **29**: 196–209.
9. Hisako Urabe, Yoko Sugawara, Mitsuo Ataka and Allan Rupprecht. 1998. Low-Frequency Raman Spectra of Lysozyme Crystals and Oriented DNA Films: Dynamics of Crystal Water. Biophys. J. **74**: 1533–1540
10. Hédoux A, Ionov R, Willart J-F, Lerbret A, Affouard F, Guinet Y, Descamps M, Prévost D, Paccou L and Danéde F, 2006. Evidence of a two-stage thermal denaturation process in lysozyme: A Raman scattering and differential scanning calorimetry investigation. J. Chem. Phys. **12**4: 014703-1-7.
11. Ionov R, He´doux A, Guinet Y, Bordat P, Lerbret A, Affouard F, Prevost D and Descamps M, 2006. Sugar bioprotective effects on thermal denaturation of lysozyme: Insights from Raman scattering experiments and molecular dynamics simulation J. Non-Cryst. Sol. **352**: 4430 –4436
12. Razumas V, Talaikyte Z, Barauskas J, Larsson K, Miezis Y and Nylander T, 1996. Effects of distearoylphosphatidylglycerol and lysozyme on the structure of the monoolein-water cubic phase: X ray diffraction and Raman scattering studies. Chem. and Phys. of Lipids. **84:** 123-138.
13. Lu J, Cow P-S and Carpenter K, 2003. Phase transitions in lysozyme solutions characterized by differential scanning calorimetry. Prog. Cryst. Growth Charact. Mat. **46**: 105-129.